\documentclass[twocolumn,prl,aps,floatfix,superscriptaddress,amssymb]{revtex4-1}
\usepackage{epsfig}
\usepackage{amsmath}

\begin{document}

\title{Spin and electronic correlations in gated graphene quantum rings }

\author{P.~Potasz}
\affiliation{Institute for Microstructural Sciences, National Research Council of Canada
, Ottawa, Canada}
\affiliation{Institute of Physics, Wroclaw University of Technology, Wroclaw, Poland}

\author{A.~D.~G\"u\c{c}l\"u}
\affiliation{Institute for Microstructural Sciences, National Research Council of Canada
, Ottawa, Canada}

\author{P.~Hawrylak}
\affiliation{Institute for Microstructural Sciences, National Research Council of Canada
, Ottawa, Canada}

\date{\today}

\begin{abstract}
We present a theory of graphene quantum rings designed to produce degenerate
shells of single particle states close to the Fermi level.  We show that
populating these shells with carriers using a gate leads to correlated ground
states with finite total electronic spin. Using a combination of tight-binding
and configuration interaction methods we  predict ground state and total spin
of the system as a function of the filling of the shell.  We show that for
smaller quantum rings, the spin polarization of the ground state at half
filling depends strongly on the size of the system, but reaches a maximum
value after reaching a critical size. 
\end{abstract}

\maketitle

\section{I. Introduction}
There is currently significant interest in developing understanding of
electronic  properties
\cite{Novoselov+Geim+04,Novoselov+Geim+05,Zhang+Tan+05,Rycerz+Tworzydlo+04,Neto+Guinea+09,Geim+Novoselov+05}
and  applications \cite{Geim+Novoselov+05,Xia+Mueller+09,Mueller+Xia+10} of graphene.
Starting with graphene as a zero gap non-magnetic material, reducing the
lateral size  and controlling the shape and character of the edges opens the
possibility of controlling the energy spectrum and hence electronic and
magnetic properties  of graphene nanostructures
\cite{Heiskanen+Manninen+08,Zhang+Chang+08,Bahamon+Pereira+09,Ezawa+06,Nakada+Fujita+96}. In
particular, the zigzag edges are responsible for degenerate energy shells at
the Fermi level
\cite{Yamamoto+Noguchi+06,Ezawa+06,Ezawa+07,Ezawa+08,Fernandez-Rossier+Palacios+07,Son+Cohen+06,Wunsch+Stauber+08},
and associated  finite spin polarization as a result of electron-electron
exchange interactions
\cite{Son+Cohen+06,Yamamoto+Noguchi+06,Ezawa+07,Ezawa+08,Fernandez-Rossier+Palacios+07,Wang+Meng+08,Wunsch+Stauber+08,Guclu+Potasz+09,Wang+Meng+09}.
However, the coupling of spin polarized zigzag edges was shown to be
anti-ferromagnetic in graphene nanoribbons \cite{Son+Cohen+06} with no net
spin polarization. In contrast, in triangular graphene quantum dots spin
polarized edges were shown to couple ferromagnetically leading to a finite
magnetic moment
\cite{Fernandez-Rossier+Palacios+07,Wang+Meng+08,Guclu+Potasz+09,Wang+Meng+09}.
The purpose of this work is to answer the question whether it is possible to
use graphene nanoribbons to build graphene nanostructures  with finite
magnetic moment. We show here that by designing a hexagonal ring from six
ribbons we obtain a quantum system with degenerate shells in the energy
spectrum. By filling these shells with additional electrons using metallic
gate we obtain maximally spin polarized ground state for the half filling of
the degenerate shell. 

Semiconductor quantum rings have been investigated by a number of groups
\cite{Aharonov+Bohm+59,Webb+Washburn+85,Bayer+Korkusinski+03,Ribeiro+Govorov+04}. The
ring geometry allows to observe quantum phenomena, e.g. persistent
current and quantum interference effects \cite{Buttiker+Imry+83}, in
particular Aharonov-Bohm (AB) oscillations \cite{Aharonov+Bohm+59}. The AB
oscillations manifest themselves as periodic oscillations in the energy
spectrum of the electronic system as a function of the number of flux quanta
entering the ring  \cite{Webb+Washburn+85}. The AB effect for a single
electron  in single lithographically defined semiconductor quantum ring
\cite{Bayer+Korkusinski+03}, hole in a type-II semiconductor dot
\cite{Ribeiro+Govorov+04} and exciton \cite{Romer+Raikh+00,Teodoro+Campo+10}
in a finite ring, was demonstrated. 

The Aharonov-Bohm conductance oscillations  were also recently observed in a
graphene ring \cite{Russo+Oostinga+08, Huefner+Molitor+09}. The electronic
properties of a single Dirac Fermion in graphene quantum rings were studied
using effective mass \cite{Abergel+Apalkov+08,Recher+Trauzettel+07} and
tight-binding methods
\cite{Recher+Trauzettel+07,Wurm+Wimmer+08,Rivera+Pereira+09}.  Valley
degeneracy in graphene was shown to be lifted by the magnetic field
\cite{Recher+Trauzettel+07, Wurm+Wimmer+08} since the magnetic field has the
opposite sign in the two valleys. 

In this work we combine the tight binding method with the configuration method
to determine the electronic and spin properties of graphene quantum rings with
zig-zag edges as a function of the number of additional electrons controlled
by the gate.  We analyze the energy spectrum of quantum rings as a function of
the width of the ring and its size. We find  degenerate electronic shells near
Fermi energy for the thinnest structures. We use configuration interaction
method to treat exactly interaction of additional  electrons in the degenerate
shell as a function of shell filling.  We determine the dependence of the spin
polarization of the ground state as a function of the filling of the
degenerate shell and the size of the structure. We show that by changing the
size of the structures we control the splitting between levels in a degenerate
shell which in turn significantly influences magnetic properties of the ground
state. The stabilization of the spin phase diagram at a critical size of the
ring is observed. 
  
This paper is organized as follows. In Section II, we introduce both the
tight-binding (TB) model for single particle levels and  configuration
interaction method for electron-electron interaction. In Section III,  we show
the method for constructing hexagonal ring structures with different width and
length. Next, in section IV we present discussion of single particle energy
spectra and explain the origin of the shell structure. In section V we analyze
the effect of electron-electron  interactions and spin properties of electrons
in degenerate shells as a function of shell filling. Finally, in section VI we
summarize obtained results.

\section{II. Model and method}
The single particle energy spectrum of $\Pi_z$ electrons in graphene quantum
rings can be obtained using the tight-binding Hamiltonian
\cite{Wallace+47}. The tight-binding model was successfully applied to  carbon
materials such as graphite, graphene, nanoribbons, nanotubes, fullerens, and
graphene quantum dots
\cite{Wallace+47,Nakada+Fujita+96,Ezawa+06,Ezawa+07,Ezawa+08,Fernandez-Rossier+Palacios+07,Yamamoto+Noguchi+06}. The
Hamiltonian in the nearest neighbors approximation can be written as 
\begin{eqnarray}
&H&=t\sum_{\left\langle i,j\right\rangle,\sigma}c^\dagger_{i\sigma}c_{j\sigma},       
\label{TB}
\end{eqnarray}
where $t$ is the hopping integral, $c^\dagger_{i\sigma}$  and $c_{i\sigma}$
are creation and annihilation operators of  electron on  $\Pi_{z}$ orbital on
site $i$ with spin $\sigma=\uparrow,\downarrow$ respectively, and
$\left\langle i,j\right\rangle$ indicate summation over
nearest-neighbours. Diagonalization of the tight-binding Hamiltonian
generates single particle energies $\epsilon_{s}$ and single particle orbitals
$|s,\sigma \rangle$.

In order to include electron-electron
interactions the many-body Hamiltonian $H_{MB}$ is written as   
\begin{eqnarray}
\nonumber
&H_{MB}&=\sum_{s,\sigma}\epsilon_{s}a^\dagger_{s\sigma}a_{s\sigma}
\\&+&\frac{1}{2}\sum_{\substack{s,p,d,f,\\\sigma,\sigma'}}\langle sp\mid \tilde{V}\mid df\rangle
a^\dagger_{s\sigma}a^\dagger_{p\sigma'}a_{d\sigma'}a_{f\sigma},       
\label{Coulomb}
\end{eqnarray}
where the first term describes single particle energies obtained from the
tight binding Hamiltonian given by Eq. (\ref{TB}) and the second term
describes interactions between particles occupying these single particle
states. By using Slater $\pi_{z}$ orbitals \cite{Ransil} we calculated
two-body Coulomb matrix elements $\langle ij\mid V\mid kl\rangle$, where
$i,j,k,l$ are the site indices.  In numerical calculations, on-site and all
scattering and exchange terms up to next nearest neighbor's are included. Few
largest Coulomb matrix elements are given in the Appendix. We use $t=-2.5$eV
for the hopping integral, and the  effective dielectric constant is set to
$\kappa=6$.

\section{III. Construction of graphene rings}

In this section we show the method for constructing a mesoscopic ring from six
graphene nanoribbons. 
\begin{figure}
\begin{center}
\epsfig{file=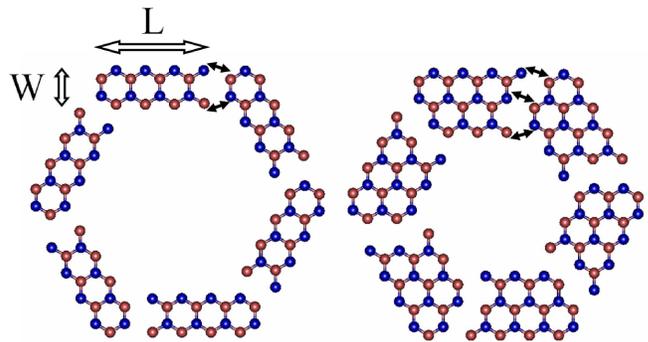,width=3.4in}
\caption{  (Color online) The method of construction of ring structures from
  smaller ribbon-like units. On the left, there are six thinnest possible
  ribbons (one benzene ring thick, what we note as $W=1$) arranged in a
  hexagonal ring structure. The length of each ribbon is given by $L=4$, the
  number of one type of atoms in one row. Each ribbon consists of $16$ atoms
  which gives a total of $96$ atoms in a ring. On the right, there are six
  ribbons with width $W=2$ (two benzene ring thick). Each of them consists of
  $21$ atoms giving a total of $126$ atoms in a ring. We create a thicker ring
  with similar length ($L=4$) but smaller antidot inside.   }
\label{fig:fig1}
\end{center}
\end{figure} 
Such a choice of building blocks helps us to understand single particle
spectrum of the ring. In Fig. \ref{fig:fig1} we show two sets of six graphene
ribbons arranged in a hexagonal ring. Each ribbon consists of two types of
atoms from the unit cell of honeycomb lattice, indicated by red (light
grey) and blue (dark grey) circles in Fig. \ref{fig:fig1}. On the left side, thinnest 
possible ribbons with one benzene ring width are shown, denoted as
$W=1$. Each of them consists of $16$ atoms (the length $L=4$,
is measured by the number of one type of atoms in the upper row), so the final
ring is built of $96$ atoms. Small black arrows indicate bonds (and hopping
integrals between nearest neighbors in a tight-binding model) between
neighboring ribbons (two arrows in the case of thinnest structures). The
number of such connecting atoms increases with increasing width as seen
on the right hand side of Fig. \ref{fig:fig1}. The thicker ribbon ($W=2$) has 
identical length to the one from the left side ($L=4$). In this case there
are three connecting atoms (three small black arrows indicate three bonds). 
The final ring is built of $126$ atoms. By connecting neighboring
ribbons with different lengths and widths we create rings with different
single particle spectra.

\section{IV. Single particle spectra}
\begin{figure}
\epsfig{file=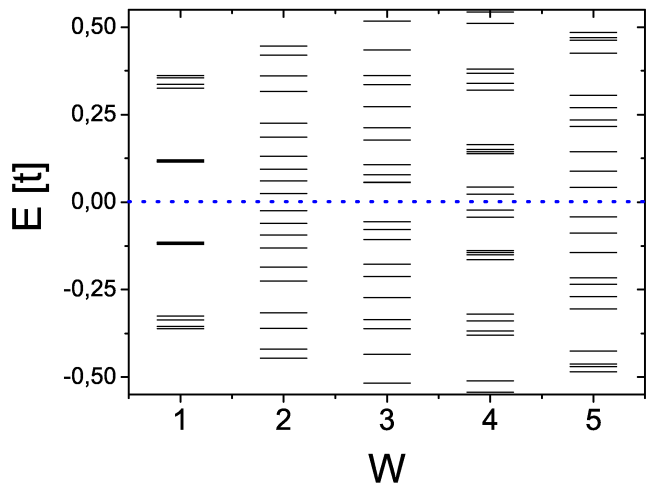,width=3.4in}
\caption{ (Color online)  Single particle spectrum near Fermi level for ring
  structures with $L=8$ and different widths. The shell structure is clearly
  observed only for the thinnest ring ($W=1$). Dotted blue (grey) line
  indicate the location of Fermi energy.   }
\label{fig:fig2}
\end{figure}

In Fig. \ref{fig:fig2} we show the  single-particle energy levels near Fermi
level obtained by diagonalizing tight-binding Hamiltonian, Eq. (\ref{TB}), for
rings with a given length ($L=8$) and different widths. The thinnest ring
($W=1$) consists of $192$ atoms. For this structure we observe nearly
degenerate shells of energy levels separated by gaps. Each shell consists of
six levels: two single and two doubly degenerate states. First shell over the
Fermi level is almost completely degenerate while in the second one the
degeneracy is slightly removed. We note that for rings with different lengths,
the gap between first and second shell is always larger then the gap at the
Fermi level. With increasing width of the ring, the spectrum changes
completely.  For the rings with width $W=2$ (270 atoms), $W=3$ (336 atoms) and
$W=5$ (432 atoms) shells are not visible. For $W=4$ (390 atoms) we observe
appearance of shells separated by gaps further from Fermi level but the
splitting between levels in these shells is much stronger in comparison to the
thinnest ring. We note that for $W\geq 2$, although we do not observe a clear
pattern of shells around the Fermi level, single shells  of six levels
separated by gaps from the rest of the spectrum appear far away from the Fermi
energy in some cases. 

\begin{figure}
\epsfig{file=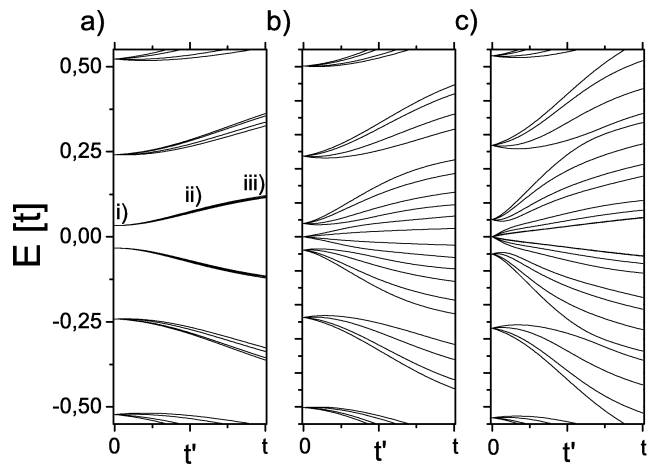,width=3.4in}
\caption{ (Color online)  The evolution of the single particle spectrum from
  six independent ribbons with $L=8$ to a hexagonal ring structure
  spectrum. $t'$ indicate hopping integrals between neighboring ribbons. a)
  For the thinnest ring ($W=1$) six fold degeneracy is slightly
  removed, preserving a shell structure. For thicker structures ((b) and (c), $W=2$
  and $W=3$ respectively) the six fold degeneracy is strongly lifted and shell
  structure is not observed.  }
\label{fig:fig3}
\end{figure}

In order to have a better understanding of the structure of the tight-binding
spectra, in Fig. \ref{fig:fig3} we show the evolution of single particle
energies from six independent ribbons to a ring  as the hopping between  the
ribbons is increased.
To achieve this, first we diagonalize tight-binding Hamiltonian matrix for a
single ribbon. We then take six such ribbons and create Hamiltonian matrix in
the basis of the eigenvectors of  six ribbons. Here, the matrix has diagonal
form. All energy levels are at least six fold degenerate. Next, using the
six ribbons basis, we write hopping integrals  corresponding to connecting
atoms between neighboring ribbons (indicated by small black arrows in
Fig. \ref{fig:fig1}). By slowly turning on the hopping integrals and
diagonalizing the Hamiltonian at every step, we can observe the evolution
of the spectrum from single particle states of six independent ribbons to a
ring. 

The hopping integrals between connecting atoms of neighboring ribbons
are indicated by $t'$ in Fig. \ref{fig:fig3}. For the thinnest ring
(Fig. \ref{fig:fig3}(a)), each ribbon consists of $32$ atoms. There are only
two connecting atoms between neighboring ribbons, giving only two  hopping
integrals $t'$ between each two ribbons in the nearest neighbors
approximation. We see that their influence is very small and six fold
degenerate states evolve into shells with a very small splitting between
levels. We note that this splitting is a bit stronger for higher energy levels
but due to large gaps between consecutive levels of single ribbon the shell
structure is still clearly observed. For the thicker structures
(Fig. \ref{fig:fig3}(b) and (c)) the evolution of the spectrum has a more
complicated behavior. For a given ring each ribbon consists of different
number of two types of atoms which give rise to zero energy
edge-states \cite{Lieb+89}. With increasing width, the number of
zero-energy states increases as well as the number of connecting atoms (and equally the
number of $t'$ hopping integrals). This causes a stronger splitting of levels
for thicker rings in comparison to the thinnest one. Thus, the
thicker ring's spectrum close to the Fermi level is due to the splitting of
zero-energy states of independent ribbons. For $W=2$ (one zero-energy state)
and $W=3$ (two zero-energy states), each ribbon consists of 45 and 56 atoms
respectively, and the evolution of their spectrum is similar. The degeneracy is
strongly lifted and no shell structure is observed. 

\begin{figure}
\epsfig{file=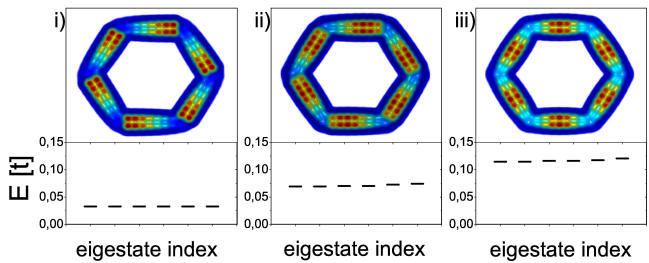,width=3.4in}
\caption{ (Color online)  Energy levels and corresponding total electronic
  densities for the first six states over the Fermi level for the thinnest
  structure with $L=8$ (192 atoms), for i) $t'=0$, ii) $t'=0.5t$, iii)
  $t'=t$. The three values of $t'$ hopping integrals are indicated in
  Fig. \ref{fig:fig3}(a).  }
\label{fig:fig4}
\end{figure}

In order to illuminate the influence of $t'$ hopping integrals on the thinnest
ring spectrum, in Fig. \ref{fig:fig4} we also show the electronic densities
for the first shell over the Fermi level for three different values of $t'$
(indicated in Fig. \ref{fig:fig3}(a)).
For $t'=0$ there are six independent ribbons and first shell is perfectly six
fold degenerate. The electronic charge density in each ribbon is larger on the
two atoms with only one bond (see Fig.\ref{fig:fig1}) and gradually decreases
along the length. For $t'=0.5t$ the total energy of the shell increases and
the degeneracy is slightly removed. Here, the highest peak of electronic
charge density is moved towards the center of each ribbon in comparison to
$t'=0$ case. Increasing $t'$ to $t$ causes  increasing of the total energy of
the shell and the highest peak of electronic charge density is now perfectly
in the middle of each arm of the ring. Thus, both electronic charge density
and energy of levels change slightly during the gradual transition of ribbons
into a hexagonal ring structure. 

\begin{figure}
\epsfig{file=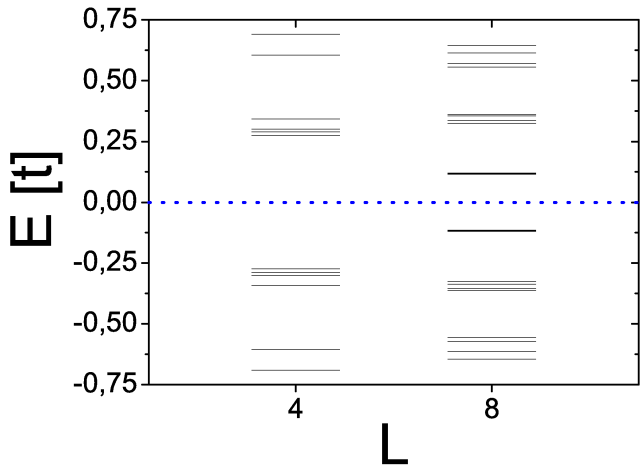,width=3.4in}
\caption{ (Color online)  Single particle spectrum near Fermi level for the
  thinnest ring structures with length $L=8$ and $L=4$. The shell structure is
  clearly observed. The splitting between levels in the first shell is smaller
  for larger structure. Dotted blue (grey) line
  indicate the location of Fermi energy.}\label{fig:fig5}
\end{figure}

We find degenerate shells near the Fermi energy  only for
the thinnest rings. In the rest of the paper, we will focus on the single and
many particle properties of these structures as a function of their
lengths. In Fig. \ref{fig:fig5} we show the low energy spectrum for two
thinnest rings with different lengths. 
We clearly see shells with six levels. The splitting of levels from the first
shell over the Fermi level is smaller for larger ring.  For ring structure
with $L=4$ the difference between the highest and the lowest energy of levels
forming the first shell is around $0.069t\eqsim 0.17$ eV. In comparison, for
ring with $L=8$ this value is around $0.006t\eqsim 0.015$ eV. Thus, we
conclude that for smaller rings single particle  energies can play important
role in the properties of many particle states while for the larger rings
interactions are expected to be more important.

\section{V. Electronic interactions in a degenerate electronic shell of a graphene quantum ring}
 
In this section, we study ground and excited states as a function of the
number of additional interacting electrons in degenerate shells of quantum
rings with different size $L$. In our calculations we assume that all states
below the Fermi level remain fully occupied. This is justified as long as
there is a sufficiently large energy gap at the Fermi level. Next, we add
extra electrons to the charge neutral system. In a first approximation we
neglect scatterings from/to the states below Fermi energy. Moreover, because
of the large energy gap between first and second shell we can neglect
scatterings to the higher energy states. This allows us to treat first shell
of the thinnest ring as an independent system which significantly reduces the
dimension of the Hilbert space. All the shells in the studied structures
consist of six levels. For a given number of extra electrons we create a basis
of all possible configurations of electrons distributed within these
states. Since the total spin of the system is conserved, we diagonalize the
many-body Hamiltonian in subspaces with total projection of spin onto z-axis
$S_{z}$.  The largest dimension of the Hilbert space is  for the half filling
($6$  electrons) and $S_{z}=0$ for which there are $400$ configurations.

First, we study the magnetic properties of the half filled shell.
\begin{figure}
\epsfig{file=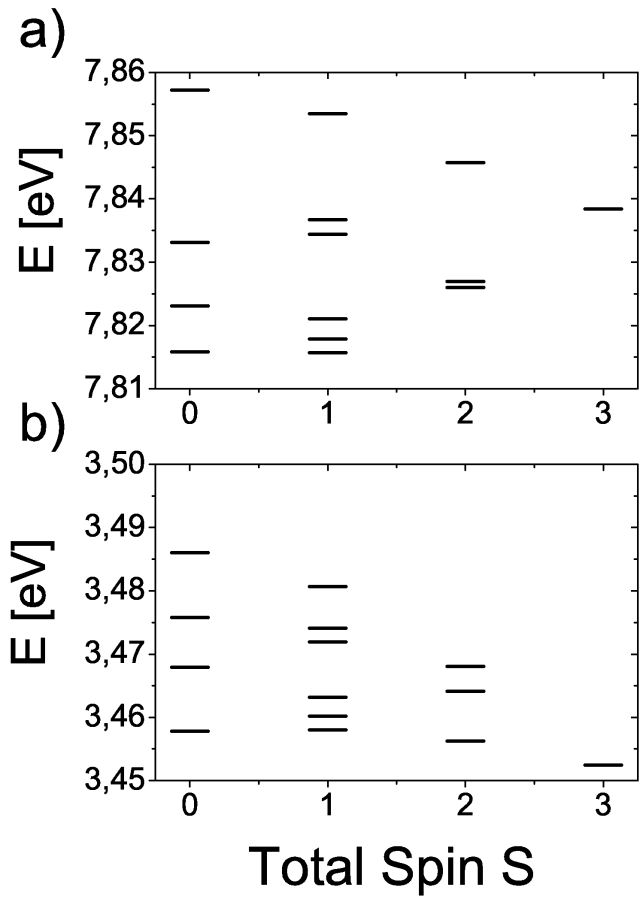,width=3.4in}
\caption{ (Color online)  The low energy spectra for the different total spin
  $S$ of half filled first shell over the Fermi energy for two thinnest rings
  with (a) $L=4$ ($96$ atoms) and (b) $L=8$ ($192$ atoms).  }
\label{fig:fig6}
\end{figure} 
Figure \ref{fig:fig6} shows the low energy spectra for the different total
spin $S$ of half filled first shell over the Fermi energy for two thinnest
rings with a) $L=4$ ($96$ atoms) and b) $L=8$ ($192$ atoms). For smaller ring
the ground state has total spin $S=1$ with a very small gap to the first
excited state with $S=0$ \cite{APPA}. The lowest states with larger total spin
have higher energies. For $192$ atoms ring the total spin of the ground state
is maximal ($S=3$). The lowest levels with different total spin have slightly
 higher energies. This can be understood in a
following way. The splitting between levels is large for smaller structures,
which is seen in Fig. \ref{fig:fig5}. For ring with $L=4$ ($96$ atoms) this
value  ($0.17$ eV) is comparable with electronic interaction terms,
e. g. $0.34$ eV for two electrons occupying the lowest state. For ring with
$L=8$ ($192$ atoms) the electron-electron interaction terms are around $0.23$
eV (for interaction between two particles on the first state), which is much
larger in comparison with single particle energy difference ($0.015$ eV). From this, we
clearly see that for ring with $L=4$ it is energetically favorable to occupy
low energy states by electrons with opposite spins. For ring with $L=8$ all
states have similar energies and due to exchange interactions the lowest
energy state is maximally spin polarized. 

\begin{figure}
\epsfig{file=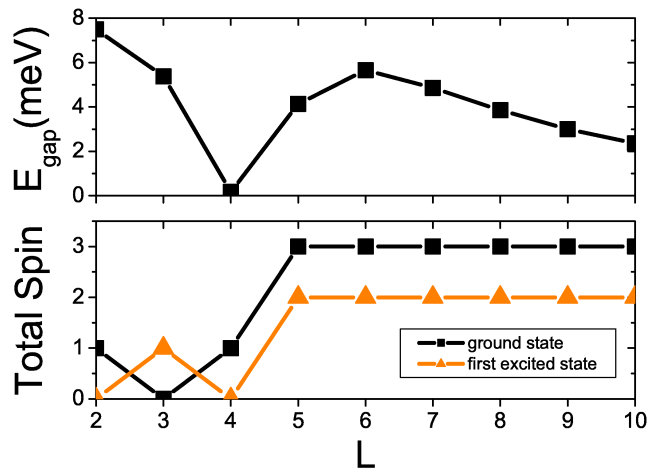,width=3.4in}
\caption{ (Color online)  Lower: Total spin of the ground and first
  excited states for the half filling of the first shell in the thinnest ring
  structures with different sizes.  Upper: Corresponding energy spin gap
  between ground and first excited states.  }\label{fig:fig7}
\end{figure}

The behavior of magnetic properties of the ground state for half filled shell
as a function of size is shown in Fig. \ref{fig:fig7}.
In this case, ground state properties can be explained as a result of the
competition between occupation of levels with smallest single particle
energies which favors opposite spin configurations, and parallel spin
configurations for which exchange interactions are maximized. For rings with
$L\geq 5$ the ground state is maximally spin polarized. Here, the splitting
between levels is relatively small and the ground state is determined by
electronic interactions. Moreover, this splitting decreases with increasing
size and this is seen in the spin gap behavior (Fig. \ref{fig:fig7}). The
largest spin gap is observed for ring with $L=6$ and decreases with increasing
$L$. For small rings the situation is more complicated. Here, the
contributions from single particle energies and interactions are
comparable. As a consequence, we observe ground states with alternating total
spin $S=1$ and $S=0$. For sufficiently large rings, $L > 5$, we observe
stabilization of the spin phase diagram. This is connected to changes of the
energy differences between levels in a shell - above a  critical size these
values are so small that they don't play a role anymore.

\begin{figure}
\epsfig{file=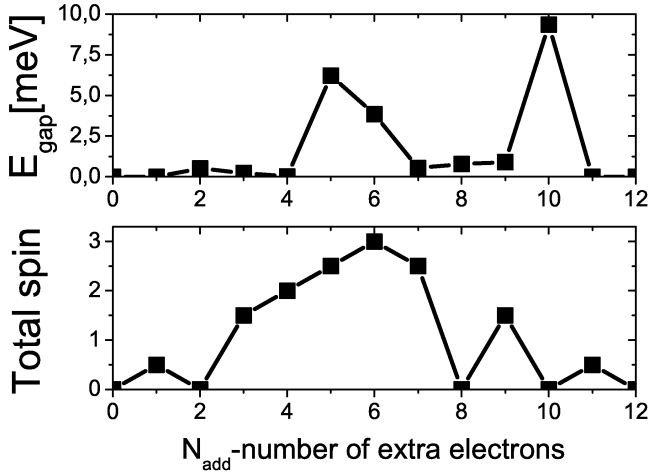,width=3.4in}
\caption{ (Color online)  The spin phase diagram for electrons occupying first
  shell of the ring structure with $L=8$ ($192$ atoms).}\label{fig:fig8}
\end{figure}

In Fig. \ref{fig:fig8} we show the phase diagram for a ring
with $L=8$ ($192$ atoms). Near the half filling the ground state is maximally
spin polarized which is related to the dominant contribution from the
short-ranged exchange interaction terms, and  charge density is symmetrically
distributed in the entire ring (see Fig. \ref{fig:fig4}). Adding or removing electrons causes
irregularities in the density distribution, and correlation effects start
becoming important. This results in an alternating spin between maximal polarization
(e.g. 3, 4, 9 extra electrons) and complete depolarization (e.g. 2, 8, 10
extra electrons) of the system.

\section{VI. Conclusions} 

We presented here a theory of graphene quantum rings designed to produce
degenerate shells of single particle states close to the Fermi level.  By
combining tight-binding and configuration interaction  methods, we analyzed
magnetic properties and electronic correlations in such structures as a
function of size and number of added electrons. For the half filling of the
degenerate shell in sufficiently large ring, maximal polarization of the
ground state is  predicted. Away from the half filling the correlation effects
appear and the ground state total spin alternates between maximal polarization
and complete depolarization.

{\it Acknowledgment}. The authors thank NRC-CNRS CRP, Canadian Institute for
Advanced Research, Institute for Microstructural Sciences, QuantumWorks
and Polish MNiSW, Grant No. N202-071-32/1513 for support.

\section{Appendix A: Calculations of Coulomb matrix elements between electrons on sites in graphene honeycomb lattice.} 
The Coulomb interaction term from eq. (\ref{Coulomb}), we can write as 
\begin{eqnarray}
\nonumber
V=\frac{1}{2}\sum_{\substack{s,p,d,f,\\\sigma,\sigma'}}\langle sp\mid V\mid df\rangle
a^\dagger_{s\sigma}a^\dagger_{p\sigma'}a_{d\sigma'}a_{f\sigma}
\\=\frac{1}{2}\sum_{\substack{s,p,d,f,\\\sigma,\sigma'}}\left[
\nonumber \sum_{\substack{i,j,k,l,\\\sigma,\sigma'}}\langle ij\mid 
\tilde{V}\mid kl\rangle A_{i}^{s}A_{j}^{p}A_{k}^{d}A_{l}^{f}
c^\dagger_{i\sigma}c^\dagger_{j\sigma'}c_{k\sigma'}c_{l\sigma}\right],       
\label{Cl}
\end{eqnarray}
where we substitute $a_{s\sigma}=\sum_{i\sigma}A_{i}^{s} c_{i\sigma}$ and
$A_{i}^{s}$ are coefficients in transformation from one basis to another. Here
the Coulomb matrix elements are defined as
\begin{eqnarray}
\nonumber
\langle ij\mid \tilde{V}\mid kl\rangle=\frac{e^{2}}{4\pi\kappa}\int\int 
d{\bf{r_{1}}}d{\bf{r_{2}}}\psi^{*}_{i}\left({\bf{r_{1}}}\right)
\psi^{*}_{j}\left({\bf{r_{2}}}\right)\times
\\\frac{1}{\mid{\bf{r_{2}}}-{\bf{r_{1}}}\mid}
\nonumber
\psi_{k}\left({\bf{r_{2}}}\right)
\psi_{l}\left({\bf{r_{1}}}\right)
\label{Cl2}
\end{eqnarray}
where $\kappa$ is an effective dielectric constant and
$\psi_{i}\left({\bf{r_{1}}}\right)$ is Slater $\pi_{z}$ orbital on a site $i$
of electron $1$, given by a function
\begin{eqnarray}
\nonumber
\psi\left({\bf{r_{1}}}\right)=\left(\frac{\xi^{5}}{32\pi}\right)^{\frac{1}{2}}
z\exp\left({\frac{-\xi{\bf{r_{1}}}}{2}}\right),       
\label{Cl3}
\end{eqnarray}
with $\xi=3.14$. Below we show selected Coulomb matrix elements for
$\kappa=1$. Numbers $1$ and $2$ and $3$ indicate electron on-site and on nearest neighbor site and on next nearest neighbor site of hexagonal lattice, respectively:
\begin{table}[tb]
\begin{displaymath}
\begin{array}{|c|c|}\hline 
\langle ij\mid \tilde{V}\mid kl\rangle & E~[eV]\\ \hline 
\langle 11\mid \tilde{V}\mid 11\rangle & 16.522\\
\langle 12\mid \tilde{V}\mid 21\rangle & 8.640\\
\langle 13\mid \tilde{V}\mid 31\rangle & 5.333\\
\langle 11\mid \tilde{V}\mid 12\rangle & 3.157\\
\langle 12\mid \tilde{V}\mid 31\rangle & 1.735\\
\langle 12\mid \tilde{V}\mid 12\rangle & 0.873\\
\langle 11\mid \tilde{V}\mid 22\rangle & 0.873\\
\langle 22\mid \tilde{V}\mid 13\rangle & 0.606\\
\hline
\end{array}
\end{displaymath}
\caption{\label{tab:xl} Selected coulomb matrix elements between electrons on sites in graphene honeycomb lattice.. Numbers $1$, $2$ and $3$ indicate electron on-site, on nearest neighbor site and on next nearest neighbor site of hexagonal lattice, respectively.}
\end{table}
\vspace*{-0.22in}


\end{document}